\pdfoutput=1

\documentclass[8.5pt,twoside,twocolumn]{article}
\usepackage[super,sort&compress,comma]{natbib} 
\usepackage{mhchem}
\usepackage{times,mathptmx}
\usepackage{sectsty}
\usepackage{balance} 

\usepackage{lastpage}
\usepackage[format=plain,justification=raggedright,singlelinecheck=false,font=small,labelfont=bf,labelsep=space]{caption} 
\usepackage{fancyhdr}
\newcommand{\eq}[1]{\begin{align} #1 \end{align}}

\newcommand{\keff}[0]{\ensuremath{k_\text{eff}}}
\newcommand{\p}[0]{\ensuremath{^\prime}}
\newcommand{\lstar}[0]{\ensuremath{\ell^*}}

\newcommand{\Nccut}[0]{\ensuremath{N_\text{c}^\text{cut}}}
\newcommand{\Ncextra}[0]{\ensuremath{N_\text{c}^\text{extra}}}

\usepackage{graphicx} 
\usepackage{amssymb}	
\usepackage{tikz}

\begin{document}

\thispagestyle{plain}
\fancypagestyle{plain}{
\renewcommand{\headrulewidth}{1pt}}
\renewcommand{\thefootnote}{\fnsymbol{footnote}}
\renewcommand\footnoterule{\vspace*{1pt}%
\hrule width 3.4in height 0.4pt \vspace*{5pt}} 
\setcounter{secnumdepth}{5}

\makeatletter 
\def\subsubsection{\@startsection{subsubsection}{3}{10pt}{-1.25ex plus -1ex minus -.1ex}{0ex plus 0ex}{\normalsize\bf}} 
\def\paragraph{\@startsection{paragraph}{4}{10pt}{-1.25ex plus -1ex minus -.1ex}{0ex plus 0ex}{\normalsize\textit}} 
\renewcommand\@biblabel[1]{#1}            
\renewcommand\@makefntext[1]%
{\noindent\makebox[0pt][r]{\@thefnmark\,}#1}
\makeatother 
\renewcommand{\figurename}{\small{Fig.}~}
\sectionfont{\large}
\subsectionfont{\normalsize} 

\fancyfoot{}
\fancyhead{}
\renewcommand{\headrulewidth}{1pt} 
\renewcommand{\footrulewidth}{1pt}
\setlength{\arrayrulewidth}{1pt}
\setlength{\columnsep}{6.5mm}
\setlength\bibsep{1pt}

\twocolumn[
  \begin{@twocolumnfalse}
\noindent\LARGE{\textbf{Stability of jammed packings I: the rigidity length scale$^\dag$}}
\vspace{0.6cm}

\noindent\large{\textbf{Carl P. Goodrich,$^{\ast}$\textit{$^{a}$} Wouter G. Ellenbroek,\textit{$^{b}$} and
Andrea J. Liu\textit{$^{a}$}}}\vspace{0.5cm}


\vspace{0.6cm}

\noindent \normalsize{In 2005, Wyart {\it et al.} [\textit{Europhys. Lett.}, 2005, \textbf{72}, 486] showed that the low frequency vibrational properties of jammed amorphous sphere packings can be understood in terms of a length scale, called $\lstar$, that diverges as the system becomes marginally unstable. Despite the tremendous success of this theory, it has been difficult to connect the counting argument that defines $\lstar$ to other length scales that diverge near the jamming transition. We present an alternate derivation of $\lstar$ based on the onset of rigidity. This phenomenological approach reveals the physical mechanism underlying the length scale and is relevant to a range of systems for which the original argument breaks down. It also allows us to present the first direct numerical measurement of $\lstar$.
}
\vspace{0.5cm}
 \end{@twocolumnfalse}
  ]
  \footnotetext{\dag~Electronic Supplementary Information (ESI) available: [details of any supplementary information available should be included here]. See DOI: 10.1039/b000000x/}
\footnotetext{\textit{$^{a}$~Department of Physics, University of Pennsylvania, Philadelphia, Pennsylvania 19104, USA; E-mail: cpgoodri@sas.upenn.edu}}
\footnotetext{\textit{$^{b}$~Department of Applied Physics and Institute for Complex Molecular Systems, Eindhoven University of Technology, P.O. Box 513, NL-5600 MB Eindhoven, The Netherlands }}

\section{Introduction}
Disordered solids exhibit many common features, including a characteristic temperature dependence of the heat capacity and thermal conductivity~\cite{Phillips:1981um} and brittle response to mechanical load.~\cite{Maloney:2007en}  A rationalization for this commonality is provided by the jamming scenario,~\cite{OHern:2003vq,Liu:2010jx} based on the behavior of packings of ideal spheres ({\it i.e.} soft frictionless spheres at zero temperature and applied stress), which exhibit a jamming transition with diverging length scales~\cite{OHern:2003vq,Silbert:2005vw,Wyart:2005wv,Wyart:2005jna,Drocco:2005ho,Ellenbroek:2006df,Ellenbroek:2009dp,Vagberg:2011fe,Goodrich:2012ck} as a function of packing fraction.  According to the jamming scenario, these diverging length scales are responsible for commonality, much as a diverging length near a critical point is responsible for universality.  

One of these length scales, the ``cutting length" $\lstar$, is directly tied to the anomalous low-frequency behavior that leads to the distinctive heat capacity and thermal conductivity of disordered solids,~\cite{Phillips:1981um} and is thus considered a cornerstone of our theoretical understanding of the jamming transition. This length arises from the so-called \emph{cutting argument} introduced by Wyart {\it et al.}~\cite{Wyart:2005wv,Wyart:2005jna} which is a counting argument that compares the number of constraints on each particle to the number of degrees of freedom in a system with free boundary conditions. Despite its importance, however, the connection between the cutting length derived by Wyart {\it et al.} and other physical length scales that diverge with the same exponent~\cite{Silbert:2005vw,Ellenbroek:2006df,Ellenbroek:2009dp} has not been understood. 


In this paper, we show that $\lstar$ is more robustly defined as a rigidity length. It is therefore relevant even for systems for which counting arguments are less useful, such as packings of frictional particles~\cite{Somfai:2007ge,Shundyak:2007ga,Henkes:2010kv} or ellipsoids,~\cite{Donev:2007go,Zeravcic:2009wo,Mailman:2009ct} or for experimental systems where it is not possible to count contacts. While this approach is applicable to these more general systems, we will use the traditionally employed soft sphere packings to motivate the rigidity length and illustrate its scaling behavior. We also show that $\lstar$ is directly related to a length scale identified by Silbert {\it et al.}~\cite{Silbert:2005vw} that arises from the longitudinal speed of sound.


\section{Model and numerical methods\label{sec:methods}}
{\it Generating mechanically stable packings.} 
We numerically generate packings of $N=4096$ frictionless disks in $d=2$ dimensions at zero temperature. Particles $i$ and $j$ interact with a harmonic, spherically symmetric, repulsive potential given by $V(r_{ij}) = \frac \epsilon 2 \left(1-r_{ij}/\sigma_{ij}\right)^2$ only if $r_{ij}<\sigma_{ij}$, where $r_{ij}$ is the center-to-center distance, $\sigma_{ij}$ is the sum of their radii and $\epsilon\equiv 1$ sets the energy scale. All lengths will be given in units of $\sigma$, the average particle diameter, and frequencies will be given in units of $\sqrt{\keff/m}$, where $\keff$ is the average effective spring constant of all overlapping particles and $m$ is the average particle mass.

Mechanically stable athermal packings were prepared with periodic boundary conditions by starting with randomly placed particles (corresponding to $T=\infty$) and then quenching the total energy to a local minimum. Energy minimization was performed using a combination of linesearch methods (L-BFGS and Conjugate gradient), Newton's method and the FIRE algorithm~\cite{Bitzek:2006bw} to maximize accuracy and efficiency. The distance to jamming is measured by the pressure, $p$, and the density of a system was adjusted until a target pressure was reached. Systems were discarded if the minimization algorithms did not converge. For reasons discussed in Sec.~\ref{sec_cutting_argument_review}, each packing was then replaced with a geometrically equivalent unstressed spring network.

The arguments we will present will concern the average number of contacts of each particle, $Z$, which approaches $2d$ in the limit of zero pressure. At positive pressure, the contact number is given for harmonic interactions by the relation $Z-2d \sim p^{1/2}$.~\cite{Durian:1995eo,OHern:2003vq}

{\it Creating a cut system.} We create a cut system by first periodically tiling the square unit cell, consistent with the periodic boundary conditions. We then remove all particles whose center is outside a box of length $L$. By first tiling the system, we are able to take cuts that are larger than the unit cell, as well as cuts that are smaller. We have checked that our results are not dependent on the choice of $N=4096$ particles per unit cell.

{\it Calculating zero modes and rigid clusters.} To calculate the vibrational modes of the unstressed spring network, we diagonalize the $dN$ by $dN$ dynamical matrix $D_{ij}^{\alpha\beta}$, which is given by the second derivative of the total energy with respect to particle positions:
\eq{	D_{ij}^{\alpha\beta} = \sum_{\left<i,j\right>}k_{ij} \frac{\partial^2 r_{ij}}{\partial r_i^\alpha \partial r_j^\beta}, }
where $r_i^\alpha$ is the $\alpha$ component of the position of particle $i$, and $k_{ij} \equiv \frac{\partial^2V(r_{ij})}{\partial^2r_{ij}}$ is the stiffness of the bond.
The eigenvectors give the polarization of each mode, and the corresponding eigenvalues are the square of the mode frequency. Note that the dynamical matrix for sphere packings, as opposed to unstressed spring networks, has an additional term that is proportional to the stress.

Using the zero modes ({\it i.e.} modes with zero eigenvalues), one can easily calculate rigid clusters directly from their definition (see Sec.~\ref{sec_cluster_argument}) . However, since only the zero modes are required to calculate rigid clusters, we use a pebble game algorithm developed by Jacobs and Thorpe~\cite{Jacobs:1995vi,Thorpe:1996uy} to understand the rigidity percolation transition in bond- and site-diluted lattices. This algorithm decomposes any network into distinct rigid clusters and can also be used to calculate the number of zero modes. We use the pebble game because its tremendous efficiency allows us to calculate rigid clusters for very large systems, although rigid clusters can always, in principle, be derived from modes of the dynamical matrix. Software for running the pebble game algorithm was obtained online at http://flexweb.asu.edu/.  

Note that zero modes, and thus rigid clusters, can be derived purely from the connectivity of the system without knowledge of the particular form of the interaction potential. Thus, our results are completely general for soft finite-ranged potentials; only the scaling between pressure and excess contact number needs to be adjusted, as described in Ref.~\cite{Liu:2010jx}, if other potentials were used. 

\section{Review of the cutting argument~\cite{Wyart:2005wv,Wyart:2005jna}\label{sec_cutting_argument_review}} 

The cutting argument~\cite{Wyart:2005wv,Wyart:2005jna} addresses the origin of the low-frequency plateau in the density of vibrational modes in jammed packings.~\cite{OHern:2003vq,Silbert:2005vw}
Consider an infinite, mechanically stable packing of soft frictionless spheres in $d$ dimensions at zero temperature and applied stress. Two spheres repel if they overlap, {\it i.e.} if their center to center distance is less than the sum of their radii, but do not otherwise interact. ``Rattler" particles that have no overlaps should be removed. Since the remaining degrees of freedom must be constrained, the average number of contacts on each particle, $Z$, must be greater than or equal to $2d$, which is precisely the jump in the contact number at the jamming transition.~\cite{OHern:2003vq,Goodrich:2012ck} 

It is instructive to study a simpler system, the ``unstressed" system, in which each repulsive interaction between pairs of particles in the system is replaced by a harmonic spring of equivalent stiffness $k$ at its equilibrium length.  The geometry of this spring network is identical to the geometry of the repulsive contacts between particles in the original system and the vibrational properties of the two systems are closely related.~\cite{Wyart:2005jna}
Now consider a square subsystem of linear size $L$ obtained by removing all the contacts between particles (or, in the language of the unstressed system, all springs) that cross the boundary between the subsystem and the rest of the infinite system.   Let the number of zero frequency modes in the cut system be $q$ and the number of these \emph{zero modes} that extend across the cut system be $q\p$.  Wyart {\it et al}.~\cite{Wyart:2005wv,Wyart:2005jna} used these modes to construct trial vibrational modes for the original infinite packing, as follows. If we restore the cut system with these $q\p$ extended zero modes back into the infinite system, the modes would no longer cost zero energy because of the contacts that connect the subsystem to the rest of the system. Trial modes are therefore created by deforming each extended zero mode sinusoidally so that the amplitudes vanish at the boundary. This deformation increases the energy of each mode to order $\omega_L^2$, where $\omega_L \sim 1/L$.

Note that if a mode is not extended, then it must be localized near the boundary, since the uncut system has no zero modes. However, the above procedure involves setting the mode amplitude to zero at the boundary, and so cannot be applied to such modes. It is therefore crucial to use only the $q\p$ extended modes to construct trial modes.

\begin{figure}[htb]
\centering
  \includegraphics[width=0.7\linewidth,angle=-90]{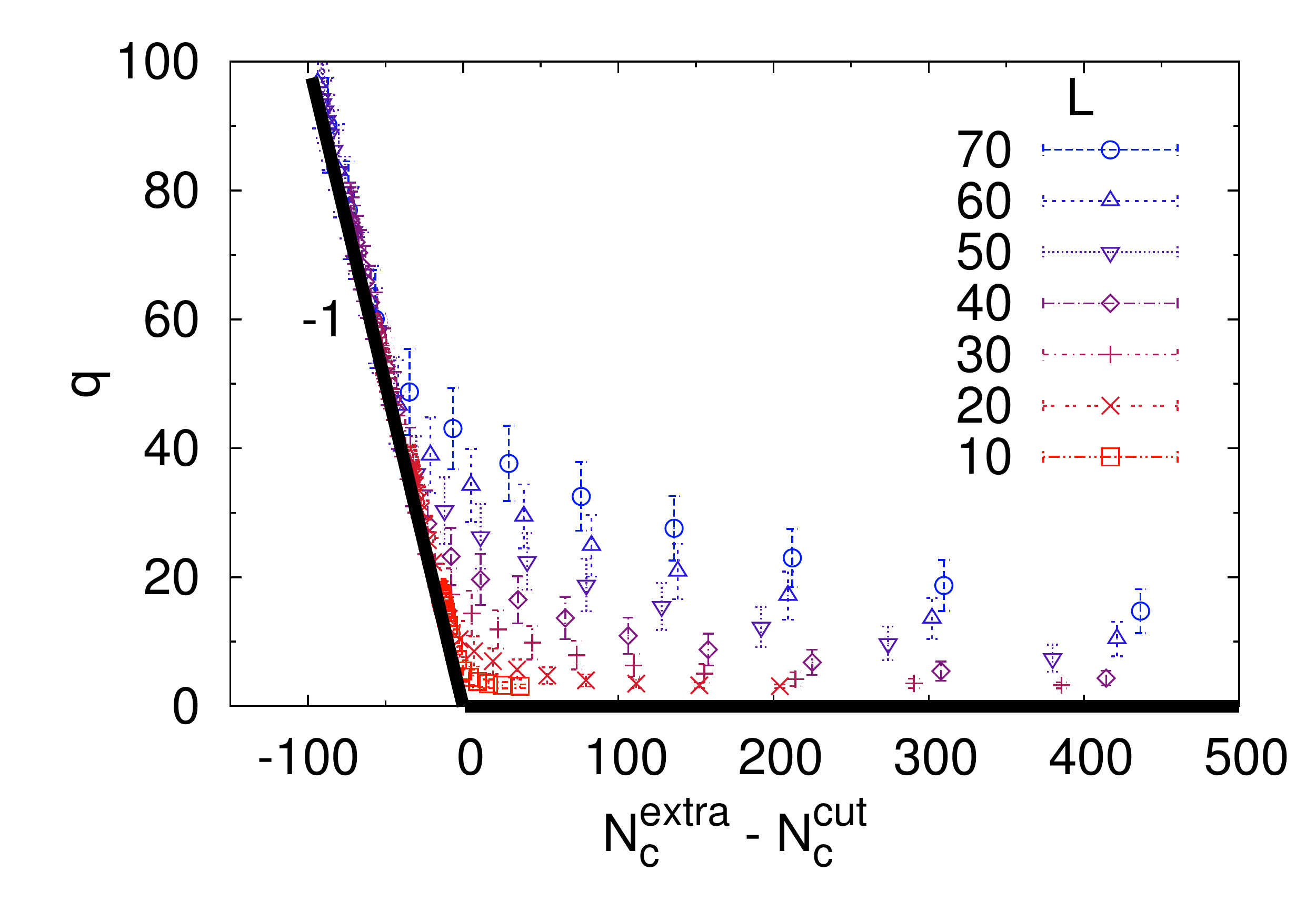}
  \caption{Number of excess zero modes as a function of the number of excess contacts after the cut. Each data point is an average of configurations at constant pressure.}
  \label{figa}
\end{figure}

The cutting argument now makes the assumption that $q\p = aq$, where $a$ is a constant independent of $L$. Before the cut, the number of extra contacts in the subsystem above the minimum required for stability is $\Ncextra \sim (Z-2d)L^d$. When the cut is made, we lose $\Nccut \sim L^{d-1}$ contacts. Naive constraint counting suggests that $q\p \sim q=\max\left(-\left(\Ncextra-\Nccut \right), 0\right)$, as shown by the solid black line in Fig.~\ref{figa}. Since $\Ncextra$ and $\Nccut$ both depend on $L$, we can define a length scale $\lstar$ by
\eq{
	\begin{array}{r l l }
		q\p\!\!\!\!\!\! &= 0 	& \mbox{if $L>\lstar$} \\
		q\p\!\!\!\!\!\! &> 0	& \mbox{if $L<\lstar$}.
	\end{array}
	 \label{lstardef}
}
The onset of zero modes is marked by $\Ncextra = \Nccut$, so
\eq{	\lstar \sim \frac{1}{Z-2d}.	\label{lstar_cuttingargument_def}}

\begin{figure*}[ht!pb]
	\centering
	\includegraphics[width=1.0\linewidth]{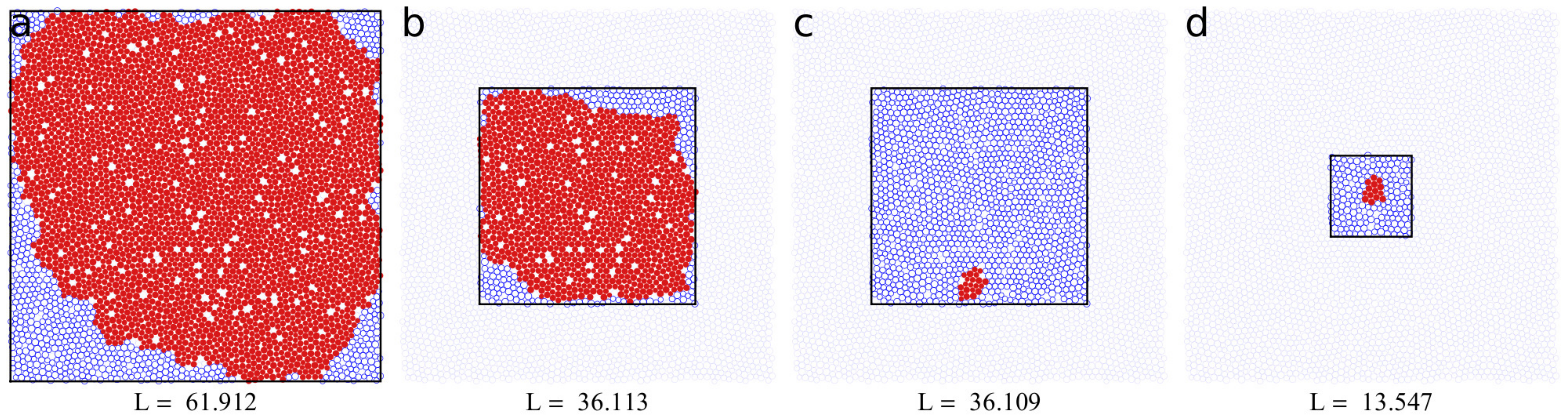}
	\caption{Subsystems cut from a $N=4096$ particle packing at a pressure $p\approx2.5\times 10^{-4}$. a) A large subsystem with $q=60$ non-trivial zero modes. Only particles circled in blue participate in the zero modes. The solid red particles form a rigid cluster.
	b) A smaller subsystem with $q=35$ zero modes. 	c) A subsystem obtained by removing one additional particle from the system in (b). This added a single additional zero mode that extends across the entire system. The largest remaining rigid cluster only contains 21 particles. The breakup of the rigid cluster from (b) to (c), and the appearance of the corresponding extended zero mode, is the phenomenon associated with the cutting length.
	d) A small system below $\lstar$ with $q=33$ zero modes.  The largest rigid cluster contains $14$ particles.
	}
	\label{figb}
\end{figure*}

The variational argument now predicts that at least $q\p/2$ of the total $L^d$ eigenmodes of the full system must have frequency less than order $\omega_L$, so the integral of the density of states from zero to $\omega_L$ must be
\eq{	\int_0^{\omega_L} \textrm{d}\omega D(\omega) \geq \frac{q\p}{2L^d}.	}
However, $D(\omega)$ is an intrinsic property of the infinite system and must be independent of $L$. Therefore, assuming no additional low frequency modes beyond those predicted by the variational argument, we can vary $L$ to back out the full density of states, as follows.

If $L>\lstar$, then $q\p=q=0$ and 
\eq{	\int_0^{\omega_L} \textrm{d}\omega D(\omega) = 0.\label{int_Domega_1}	}
For $L<\lstar$, we can write $q\p/2= a(\Nccut - \Ncextra)/2 \sim L^d\left(\omega_L - 1/\lstar\right)$, 
which leads to 
\eq{	\int_0^{\omega_L} \textrm{d}\omega D(\omega) \sim \omega_L - 1/\lstar.\label{int_Domega_2}	}
Eqns~\eqref{int_Domega_1} and \eqref{int_Domega_2} imply that 
\eq{	D(\omega) = \left\{ 
\begin{array}{l l}
	0 & \quad \mbox{if $\omega < \omega^*$}\\
	\text{const.} & \quad \mbox{if $\omega > \omega^*$,}\\ 
\end{array} \right. }
where $\omega^* \equiv 1/\lstar \sim Z-2d$ defines a frequency scale. Note that while $\lstar$ is potential independent, the units of frequency, and thus $\omega^*$, depend on potential.~\cite{Liu:2010jx} This argument predicts that the density of states has a plateau that extends down to zero frequency at the jamming transition, where $Z-2d=0$. Above the jamming transition, when $Z-2d>0$, the plateau extends down to a frequency $\omega^*$ before vanishing. This agrees well with numerical results on the unstressed system.~\cite{Wyart:2005jna}  Note the importance of the length scale $\lstar$, which defines the frequency scale $\omega^*$ and is responsible for the excess low frequency modes.

\section{Too many zero modes\label{sec_lstar_problems}}

In the cutting argument, the length scale $\lstar$ is defined as the size of a cut region, $L$, where the number of extended zero modes, $q\p$, first vanishes (eqn~\eqref{lstardef}). The argument then assumes that this coincides with the disappearance of all nontrivial zero modes, $q$, which is assumed to occure when the cut system is isostatic ({\it i.e.} when $\Nccut = \Ncextra$). Wyart {\it et al.} showed~\cite{Wyart:2005jna} numerically that this is true when $Z=2d$, but they do not provide such evidence for over-constrained systems. 

Fig.~\ref{figb}a shows a system that remains over-constrained after the cut ($\Ncextra>\Nccut$). The cutting argument would assert that the only zero modes are the trivial global translations and rotations, but we find that there are in fact $60$ non-trivial zero modes. This is generalized in Fig.~\ref{figa}, which shows that $q>0$ for all cut sizes $L$ and values of $\Ncextra-\Nccut$. Clearly, one cannot use the onset of zero modes to determine $\lstar$.

However, note that the zero modes in Fig.~\ref{figb}a exist only around the boundary (the particles depicted by blue circles), while \emph{none} of the non-trivial zero modes extend into the region of solid red particles. Since these zero modes are not fully extended, the system is above the cutting length.  As noted by Wyart et al.,~\cite{Wyart:2005wv} the scaling of the cutting argument would still be robust if the number of these excess boundary zero modes scales as $L^{d-1}$.  However, as can be seen in Fig.~\ref{figb}a, these modes penetrate a non-negligible distance into the bulk of the system and so this scaling is not obvious.


%
%
%

\section{Cluster argument\label{sec_cluster_argument}}
We now reformulate the cutting argument in a way that does not rely on the \emph{total} number of zero modes but is specifically designed to identify the onset of \emph{extended} zero modes, which are the ones needed to obtain $\omega^*$.
The mathematics will be similar to that in the cutting argument, but the setup and interpretation will be different. We will first introduce the idea of rigid clusters and illustrate the associated phenomenon that identifies the cutting length. We will then provide a rigorous derivation of the scaling of $\lstar$ in jammed packings.

Our argument is motivated by the simple fact that if none of the zero modes are extended, then by definition there must be a cluster of central particles that these modes do not reach. Since this cluster does not participate in any zero modes, any deformation to the cluster increases its energy. Thus, such clusters have a finite bulk modulus and we will refer to them as being \emph{rigid}. The solid red particles in Fig.~\ref{figb}a are an example of a rigid cluster. 
To be precise, a rigid cluster is defined as a group of particles (within an infinite $d$ dimensional system with average contact number $Z$) such that, if all other particles were removed, the only zero modes in the unstressed system would be those associated with global translation and rotation. This is purely a geometrical definition and is independent of potential. 

Fig.~\ref{figb}b-d shows the same system as Fig.~\ref{figb}a, except with progressively smaller cut regions. Fig.~\ref{figb}a and b are both dominated by a rigid cluster that covers approximately 84\% of the cut region. However, while the cut region in Fig.~\ref{figb}c differs from that in Fig.~\ref{figb}b by only a single particle, it has no rigid cluster larger than 21 particles (it is comprised of many small rigid clusters, the largest of which is shown in red). Apparently, the removal of a single particle introduced a zero mode that extends throughout the system and is precisely the type needed by the variational argument of Wyart {\it et al.}~\cite{Wyart:2005wv,Wyart:2005jna}

This sudden breakup of the rigid cluster, which coincides with the onset of extended zero modes, is a non-trivial phenomenon that marks the length scale $\lstar$. We will now provide a formal derivation of this phenomenon, which leads to a clear physical definition of $\lstar$ and allows us to derive its scaling.


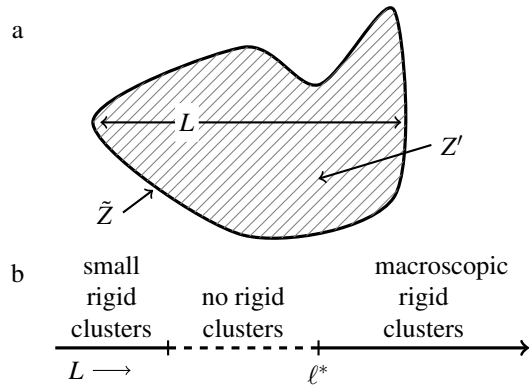
\begin{figure}[tpb]
	\[\begin{tikzpicture}
	\draw (-1.,1.2) node{a};
	\draw[very thick] plot [smooth cycle] coordinates {(0,0) (2,1) (3, 0.5) (4, 1.5) (4,-1) (2,-1.5)};
	\begin{scope}
	\clip plot [smooth cycle] coordinates {(0,0) (2,1) (3, 0.5) (4, 1.5) (4,-1) (2,-1.5)};
	\foreach \x in {-2,-1.85,...,4}
	{
		\draw[gray] (\x,-2) -- (\x+4,2);
	}
	\draw[<->,thick] (0.1,0) -- (4.1,0);
	\filldraw[white] (1.1,0.25) rectangle (1.4,-0.25);
	\draw (1.25,0) node{$L$};
	\end{scope}
	\draw[thick,->] (4.5,-0.3) node[right] {$Z^\prime$} -- (3,-0.75) ;
	\draw[thick,->] (0.4,-1.2) node[left] {$\tilde Z$} -- (0.8,-0.9) ;	
	\draw (-1.,-2) node{b};
	\draw[very thick] (-0.5,-3) --  node[above]{\text{\parbox{37pt}{\centering small rigid clusters}}} (1,-3) ;
	\draw[thick] (1,-2.9) -- (1,-3.1);
	\draw[very thick,dashed] (1,-3)  -- node[above]{\text{\parbox{37pt}{\centering no rigid clusters}}}(3,-3);
	\draw[thick] (3,-2.9) -- (3,-3.1) node[below] {$\ell^*$};
	\draw[very thick,->] (3,-3) --  node[above]{\text{\parbox{37pt}{\centering macroscopic rigid clusters}}}(5.8,-3);
	\draw[->](0,-3.3) node[left]{$L$} -- (0.5,-3.3);
	\end{tikzpicture}\]
	\caption{a) An arbitrary surface (solid black line) of size $L$ and an enclosed rigid cluster (stripes). The rigid cluster has an average contact number of $Z\p$ in the bulk and $\tilde{Z}$ at the boundary. As $L$ becomes large, fluctuations in $Z\p$ and $\tilde{Z}$ vanish. b) Possible values of $L$ such that a rigid cluster fits within the surface. Rigid clusters can either be small or larger than some minimum value. This minimum value defines $\lstar$.}
	\label{fig_clstr_ex}
\end{figure}

Consider an arbitrary $d-1$ dimensional closed surface with characteristic size $L$ (for example, the solid black curve in Fig.~\ref{fig_clstr_ex}a). We will begin by asking whether or not it is \emph{possible} for all the particles within this surface to form a single rigid cluster.
For the cluster to be rigid, it must satisfy
\eq{	N_\text{c} - dN \ge -\frac 12 d(d+1) \label{clstr_ineq_1}}
where $N$ and $N_\text{c}$ are the number of particles and contacts in the cluster, respectively, and $\frac 12 d(d+1)$ is the number of global translations and rotations. This is a necessary but not sufficient condition for rigidity.
We can write $N_\text{c}$ as
\eq{	N_\text{c} = \frac 12 Z\p(N - N_\text{bndry}) + \frac 12 \tilde{Z}N_\text{bndry}	,	}
where $\tilde{Z}$ is the contact number of the $N_\text{bndry}$ particles on the boundary and $Z\p$ is the contact number of the particles not on the boundary (see Fig.~\ref{fig_clstr_ex}a). 
We can also define the positive constants $a$ and $b$ such that $N = 2aL^d$ and $N_\text{bndry} = 2bL^{d-1+\gamma}$, where $\gamma\ge 0$ depends on the shape of the surface, with $\gamma =0$ for non-fractal shapes.\footnote[3]{For now, we place no restrictions on the fractal dimension of the shape.} 
For shapes that have multiple characteristic lengths, {\it e.g.} a long rectangle, the choice of which length to identify as $L$ is irrelevant as it only leads to a change in the constants $a$ and $b$. For concreteness, we will always take $L$ to be the radius of gyration. 

Eqn~\eqref{clstr_ineq_1} now becomes
\eq{	aL^{d-1+\gamma} \left( (Z\p-2d)L^{1-\gamma} - c \right) \ge -\frac 12 d(d+1), \label{clstr_ineq_2}	}
where $c=\frac ba(Z\p-\tilde{Z})>0$. Eqn~\eqref{clstr_ineq_2} is trivially satisfied if $(Z\p-2d)L^{1-\gamma}-c > 0$, which implies
\eq{	L> L_\text{min}(Z\p,c,\gamma) \equiv \left(\frac{c}{Z\p-2d}\right)^{1/(1-\gamma)}. \label{L_bound}	} 
We will refer to clusters that satisfy eqn~\eqref{L_bound} as \emph{macroscopic} clusters. However, it is also possible for $(Z\p-2d)L^{1-\gamma}-c < 0$, provided $L$ is very small, because the right hand side of eqn~\eqref{clstr_ineq_2} is small and negative. 

It follows that it is only possible for the particles in our arbitrary surface to form a rigid cluster if the cluster is either very small or larger than $L_\text{min}$; rigid clusters of intermediate sizes cannot exist! 
Rigid clusters cannot exist below $L_\text{min}$ because the balance between the over constrained bulk and the under constrained boundary shifts towards the boundary as the cluster size decreases.  On the other hand, if a cluster is sufficiently small, then it can be rigid, as can be seen from the following constraint count for a triangular cluster of three particles.  For this cluster, there are six degrees of freedom, three constraints and three zero modes.  Because the three zero modes correspond to rigid translation in two directions and rigid rotation, they do not destroy the rigidity of the cluster.

Note that if $L$ is large, then fluctuations in $Z\p$ and $c$ vanish and $Z\p = Z$. $L_\text{min}$ is thus constant for all translations and rotations of the surface and is independent of $L$, depending only on the actual shape of the surface. 

Given our arbitrary shape parameterized by $c$ and $\gamma$, and the infinite packing parameterized by $Z$, $L_\text{min}(Z,c,\gamma)$ is the minimum possible size of any macroscopic rigid cluster in the $Z - 2d \ll 1$ limit.
However, we wish to find the minimum size of any rigid cluster \emph{regardless of shape}, which we do by finding $c^*$ and $\gamma^*$ that minimize $L_\text{min}$ and defining $\lstar \equiv L_\text{min}(Z,c^*,\gamma^*)$.
In the limit $Z \rightarrow 2d$, we immediately see that $\gamma^* = 0$ and
\eq{	\lstar = \frac {c^*}{Z-2d}. \label{lstar_scaling_prediction}}
As depicted in Fig.~\ref{fig_clstr_ex}b, we are left with the result that rigid clusters must either be very small or larger than $\lstar$, which we now interpret as a rigidity length.

\subsection{Estimating an upper bound}
We will now derive an upper bound for the magnitude of $\lstar$ in the $Z\rightarrow 2d$ limit. Since $c \sim LN_\text{bndry}/N$, $c$ is minimized when the shape is a $d$ dimensional hypersphere. We can approximate $N$ and $N_\text{bndry}$ to be $N \approx \phi V_d^L$ and $N_\text{bndry} \approx \phi S_{d-1}^L$, where $\phi$ is the packing fraction and $V_d^L$ and $S_{d-1}^L$ are the volume and surface area of a $d$ dimensional hypersphere with radius of gyration $L$. Using $S_{d-1}^L/V_d^L = w_dd/L$, where $w_d$ is the ratio of the radius of gyration of a hypersphere to its radius,\footnote[4]{$w_2 = \sqrt{1/2}$ and $w_3=\sqrt{3/5}$}
the $Z\rightarrow 2d$ limit of $\lstar$ becomes
\eq{	\lstar \approx \frac{w_dd (2d-\tilde{Z})}{Z-2d}. \label{lstar_derivation}	}
Eqn~\eqref{lstar_derivation} is a quantitative derivation of $\lstar$ as a function of $Z$ that depends only on the value of $\tilde{Z}$, the average contact number at the boundary. 

We put an upper bound on $\lstar$ by obtaining a lower bound for $\tilde{Z}$. Note that any particle at the boundary of the rigid cluster cannot have $d$ or fewer contacts. Removing such a particle would remove $d$ degrees of freedom and at most $d$ constraints, and so the rigidity of the rest of the cluster would not be affected. Thus, $\tilde{Z} \geq d+1$ and 
\eq{	\lstar \leq \frac {w_dd(d-1)}{Z-2d}. \label{lstar_upper_bound}}
%

\begin{figure}[htpb]
	\centering
	\includegraphics[width=0.8\linewidth]{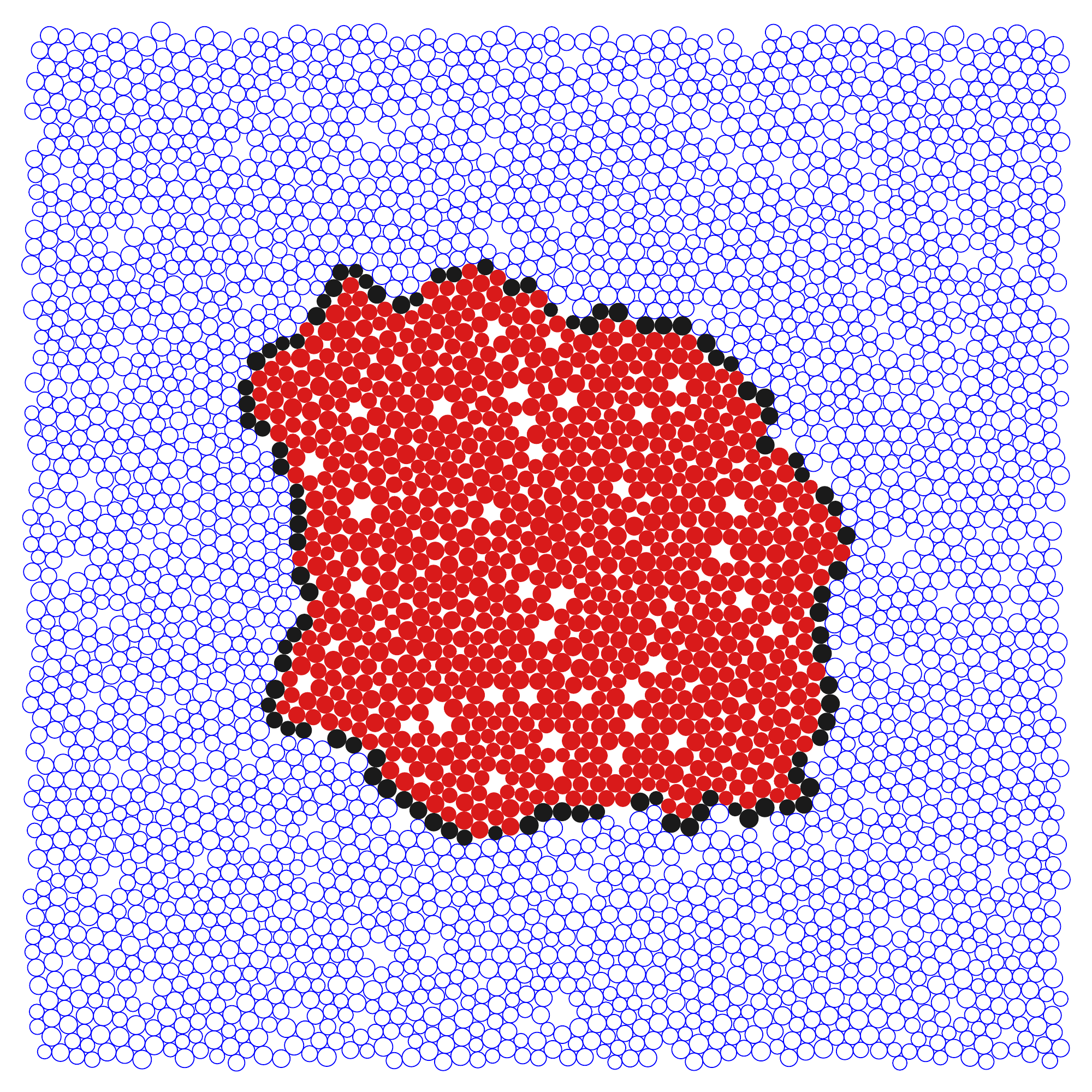}
	\caption{The smallest macroscopic rigid cluster for the system depicted in Fig.~\ref{figb}. The rigidity of the cluster formed by the solid red and black particles is destroyed if any of the black boundary particles are removed. None of the red particles make physical contact with the blue particles (which are not in the rigid cluster) and are not considered part of the boundary. The rigidity length, which is defined as the radius of gyration of the cluster, is $\lstar = 12.8$ (in units of the average particle diameter). }
	\label{fig:min_cluster}
\end{figure}

\subsection{Numerical verification\label{sec:numerical_verification}}

We will now use the cluster argument to calculate $\lstar$ numerically. Note that the rigid cluster in Fig.~\ref{figb}b is not necessarily the \emph{smallest} rigid cluster. The cluster breaks apart when the particle closest to the edge is removed (Fig.~\ref{figb}c), but it is possible that other particles at the edge of the rigid cluster can be removed without destroying the rigidity. The minimum rigid cluster that defines $\lstar$ has the property that rigidity is lost if any boundary particle is removed.

We calculate $\lstar$ by taking a large cut system (see Sec.~\ref{sec:methods}) and finding the smallest macroscopic rigid cluster.
To do this, we remove a particle that is randomly chosen from the boundary and decompose the remaining particles into rigid clusters. If there is no longer a macroscopic rigid cluster, then the boundary particle was necessary for rigidity and is put back. If the rigid cluster remains then the particle was not necessary for rigidity and we do not replace it. This process is repeated with another randomly chosen boundary particle until all the particles at the boundary of the rigid cluster are deemed necessary for rigidity. See the Electronic Supplementary Information for a video that demonstrates this process. The resulting rigid cluster ({\it e.g.} see Fig.~\ref{fig:min_cluster}) cannot be made any smaller and so its radius of gyration measures $\lstar$.

\begin{figure}[htpb]
	\centering
	\includegraphics[width=1\linewidth]{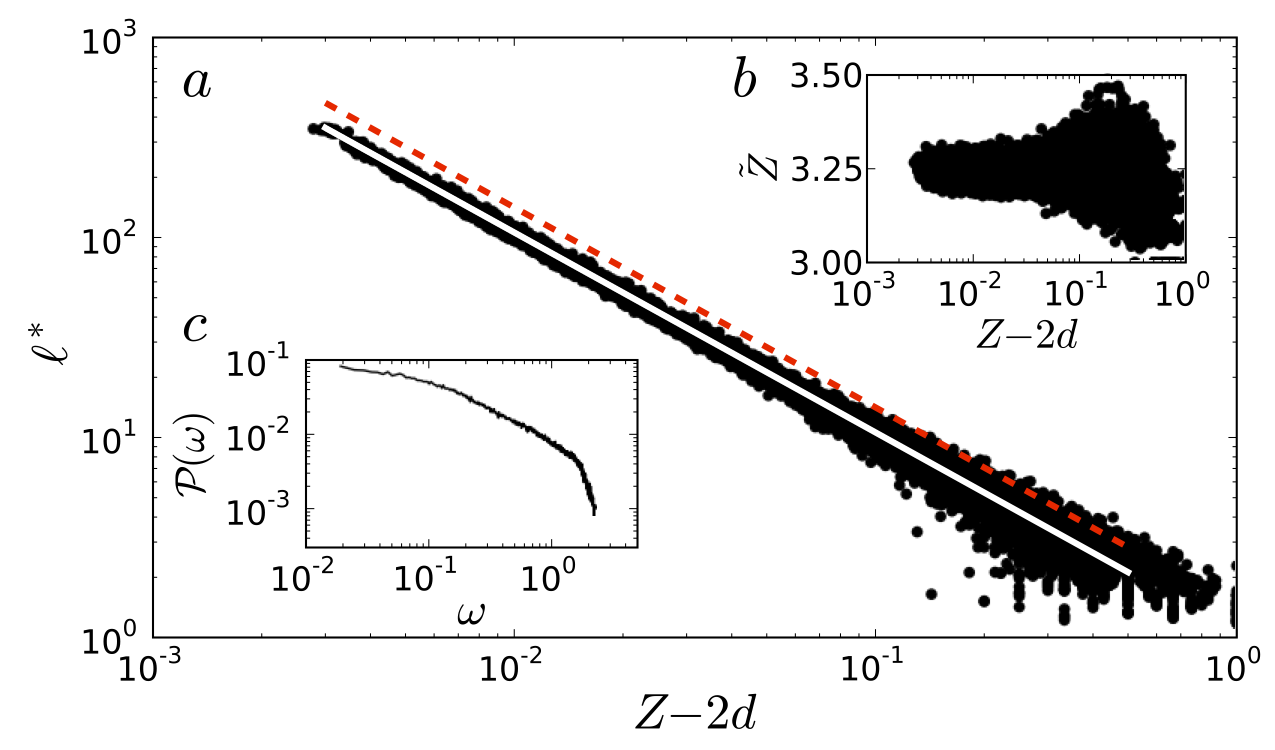}
	\caption{a) $\lstar$ as a function of $Z-2d$, measured for individual systems as described in the text. b) $\tilde{Z}\approx 3.25$ in the limit $Z\rightarrow 2d$, close to the predicted bound. The solid white line in a) is the quantitative prediction of eqn~\eqref{lstar_derivation} using $\tilde{Z} = 3.25$, while the dashed red line is the upper bound obtained from $\tilde Z = 3$. c) The projection, $\mathcal{P}(\omega)$, of the single extended zero mode just below $\lstar$ onto the modes of frequency $\omega$ in the uncut system, averaged over many realizations.}
	\label{fig_lstar_scatter}
\end{figure}

Fig.~\ref{fig_lstar_scatter}a shows that $\lstar$ diverges as $(Z-2d)^{-1}$, consistent with the cutting argument and our reformulation. In the small $Z-2d$ limit, $\lstar$ is just below the theoretical upper bound of eqn~\eqref{lstar_upper_bound} (red dashed line). Fig.~\ref{fig_lstar_scatter}b shows that $\tilde Z$, the contact number of boundary particles, is approximately $3.25$ as $Z \rightarrow 2d$, slightly above the lower bound of $3$. The solid white line in Fig.~\ref{fig_lstar_scatter}a shows the quantitative prediction from eqn~\eqref{lstar_derivation} using $\tilde{Z} = 3.25$, which agrees extremely well with the data.

According to ref.~\cite{Wyart:2005wv}, the extended zero modes of the cut system should be good trial modes for the low frequency modes of the system with periodic boundaries. Consider a system just below $\lstar$ so that there is only one extended zero mode. The global translations and rotations, as well as the boundary zero modes, can be projected out of the set of zero modes by comparing them to the modes of the system just above $\lstar$. Fig.~\ref{fig_lstar_scatter}c shows the projection of that single extended zero mode onto the $dN$ modes of the full uncut system as a function of the frequency of the uncut modes. This mode projects most strongly onto the lowest frequency modes, implying that it is, in fact, a good trial mode from which to extract the low frequency behavior, as assumed.~\cite{Wyart:2005wv} Along with the first direct measurement of $\lstar$, our results provide the first numerical verification that the trial modes of the variational argument are highly related to the low frequency modes of the periodic system. 

\subsection{Advantages of the cluster argument over the counting argument}
Along with adequately dealing with the excess zero modes in Fig.~\ref{figa}, the cluster argument has a few additional advantages.
In it, $\lstar$ is defined as the smallest rigid cluster, regardless of shape, whereas the cutting argument has to specify a flat cut. This is a potential issue because the value of $\lstar$ is sensitive to the shape of the cut.  For example, if one were to consider a shape with a non-trivial fractal dimension, then $\Nccut$ would no longer scale as $L^{d-1}$, resulting in a length with entirely different scaling. Wyart {\it et al.}~\cite{Wyart:2005wv,Wyart:2005jna} argue that a flat cut is a reasonable choice for the purposes of their variational argument, but a physical length scale with relevance beyond the variational argument should be more naturally defined. The cluster argument not only provides such a physical definition, it explains unambiguously why a flat, non-fractal cut was the correct choice in the cutting argument. 

Furthermore, defining $\lstar$ in terms of the number of zero modes can be problematic. For example, rattlers must be removed and internal degrees of freedom like particle rotations must be suppressed. For packings of ellipsoidal particles, to take one example, the choice of degrees of freedom is critical.  Jammed packings of ellipsoids lie below isostaticity~\cite{Donev:2007go} and their unstressed counterparts can have an extensive number of extended zero modes. Despite this, when the aspect ratios of the ellipsoids are small, there is a band of modes similar to those for spheres, with a density of states that exhibits a plateau above $\omega^*\sim Z - 2d$.~\cite{Zeravcic:2009wo} One would thus expect a length scale $\lstar \sim 1/\omega^*$, but constraint counting of the cutting argument does not predict this. While the cluster argument also relies on zero modes and thus cannot be applied directly in this case, the intuition that $\lstar$ is a rigidity length scale should carry over. Packings of ellipsoids can have zero modes and still be rigid, and the cluster argument would predict that there is a length scale below which a packing with free boundaries loses its rigidity.

Experimental systems present a similar challenge because the contact network is often difficult to determine. However, our result that $\lstar$ marks a rigidity transition suggests that the elastic properties of a system could be used to measure $\lstar$. Such a measurement should be experimentally tractable, would not require knowledge of the vibrational properties, and would not require specification of the degrees of freedom of the system.

\subsection{Additional comments}
As in the cutting argument, the cluster argument assumes that spatial fluctuations in $Z$ are negligible.
Wyart {\it et al.} argue~\cite{Wyart:2005jna} that fluctuations in $Z$ are negligible in $d>2$ dimensions, and that the condition of local force balance suppresses such fluctuations even in $d=2$ in jammed packings. We have applied our procedure from sec.~\ref{sec:numerical_verification} to bond-diluted hexagonal lattices where these fluctuations are not suppressed. Although these systems display a global rigidity transition~\cite{Jacobs:1995vi,Thorpe:1996uy} when they have periodic boundary conditions, they do not exhibit an abrupt loss of rigidity at some length scale that could be interpreted as $\lstar$ when they have free boundary conditions. It remains to be seen if $\lstar$ exists in this sense for bond-diluted 3 dimensional lattices.

Finally, our result that rigid clusters cannot exist on length scales below $\lstar$ appears to be consistent with results of Tighe,~\cite{Tighe:2012gm} as well as that of D\"uring {\it et al.},~\cite{During:2012bs} for floppy networks below isostaticity.  There, they find that clusters with free boundaries replaced by pinned boundaries cannot be rigid for length scales above $1/|Z-2d|$. The use of pinning boundary particles has also been used by Mailman and Chakraborty~\cite{Mailman:2011hz} to calculate a point-to-set correlation length above the transition that appears to scale as $\lstar$.

\section{Discussion\label{sec_discussion}}
We have reformulated the cutting argument in terms of rigidity instead of constraint counting. Networks derived form sphere packings can only be rigid when they have free boundaries if they are larger than a characteristic length $\lstar$, which diverges at the jamming transition. Systems just smaller than this rigidity length exhibit extended zero modes that are highly correlated with the anomalous low-frequency modes of the periodic systems, confirming the variational argument of Wyart {\it et al.}~\cite{Wyart:2005wv,Wyart:2005jna} In contrast to the original counting argument, the generalized definition of $\lstar$ does not depend on the nature of an arbitrary cut. The insight that $\lstar$ marks a rigidity transition extends the relevance of the length to systems where constraint counting is either non-trivial (such as packings with internal degrees of freedom) or not practical (such as experimental systems where determining contacts is often difficult).


The new rigidity interpretation of $\lstar$ makes it transparently clear that the cutting length $\lstar$ is equivalent to the length scale $\ell_\text{L}$, identified by Silbert {\it et al.}~\cite{Silbert:2005vw}  For systems with periodic boundaries, the anomalous modes derived from the zero modes swamp out sound modes at frequencies above $\omega^*$.  Thus, the minimum wavelength of longitudinal sound that can be observed in the system is $\ell_\text{L}=c_\text{L}/\omega^*$, where $c_\text{L}=\sqrt{B/ \rho}$ is the longitudinal speed of sound, $B \sim (Z-2d)^0$ is the bulk modulus, and $\rho$ is the mass density of the system.  

For systems with free boundaries that are smaller than $\lstar$, rigid clusters cannot exist so the bulk modulus and speed of sound vanish. The minimum wavelength of longitudinal sound that can be supported is therefore given by the minimum macroscopic cluster size, $\lstar$. From the scalings of $B$ and $\omega^*$, we see that $\ell_\text{L} \sim  \left(Z - 2d \right)^{-1} \sim \lstar$. Our definition of $\lstar$ implies that the two length scales not only have the same scaling but have the same physical meaning.  

Silbert {\it et al.} also identified a second smaller length scale $\ell_\text{T}$ from the transverse speed of sound, which depends on the shear modulus.  For systems with free boundaries to be rigid, they must support both longitudinal and transverse sound, and so while our reasoning applies to both $\ell_\text{L}$ and $\ell_\text{T}$, $\lstar$ should be the larger of the two, so that the condition for rigidity for a cluster of size $L$ is $L \gtrsim \lstar=\ell_\text{L}$.  Note that systems with periodic boundary conditions of size $L \gg \ell_\text{T}$ are stable to infinitesimal deformations of the shape of the boundary.~\cite{Schoenholz:2013jv,DagoisBohy:2012dh}


Ideal sphere packings have the special property that the number of contacts in a packing with periodic boundary conditions is exactly isostatic at the jamming transition in the thermodynamic limit.~\cite{OHern:2003vq,Goodrich:2012ck}  Here, we have shown that the number of contacts in such a system with \emph{free} boundary conditions is exactly isostatic (eqn~\eqref{clstr_ineq_1} is satisfied with a strict equality) in the cluster of size $\lstar$.   This simplicity makes ideal sphere packings a uniquely powerful model for exploring the marginally jammed state.

We thank Bryan Chen, Vitaliy Kapko, Tom Lubensky, Sidney Nagel, Daniel Sussman, and Michael Thorpe for important discussions. This research was supported by the U.S. Department of Energy, Office of Basic Energy Sciences, Division of Materials Sciences and Engineering under Award DE-FG02-05ER46199 (AJL), and by the Netherlands Organization for Scientific Research (NWO) through a Veni grant (WGE).  CPG was supported by the NSF through a Graduate Research Fellowship.


\balance

%

\footnotesize{
\providecommand*{\mcitethebibliography}{\thebibliography}
\csname @ifundefined\endcsname{endmcitethebibliography}
{\let\endmcitethebibliography\endthebibliography}{}

}

\end{document}